\documentclass[prb,amsmath,preprintnumbers,nofootinbib]{revtex4}
\usepackage{amsmath,amsfonts,amssymb,bm}
\usepackage{graphicx}

\DeclareMathOperator{\trace}{Tr}
\DeclareMathOperator{\smalltrace}{tr}
\begin{document}

\title{Gaussian phase-space representation of Fermion dynamics; Beyond the time-dependent-Hartree-Fock approximation}
\author{Saar Rahav}
\author{Shaul Mukamel}
\address{Department of Chemistry, University of California, Irvine, CA 92697}

\begin{abstract}
 A Gaussian operator representation for the many body density matrix of fermionic systems, developed by Corney and Drummond [{\em Phys. Rev. Lett}, {\bf 93}, 260401 (2004)], is used to derive approximate decoupling schemes for their dynamics. In this approach the reduced single electron density matrix elements serve as stochastic variables which satisfy an exact Fokker-Planck equation. The number of variables scales as $\sim N^2$ rather than $\sim \exp (N)$ with the basis set size, and
the time dependent Hartree Fock approximation (TDHF) is recovered in the ``classical'' limit.
 An approximate closed set of equations of motion for the one and two-particle reduced density matrices, provides a direct generalization of the TDHF. 
\end{abstract}

\pacs{03.65.Yz,71.10.Ca,31.15.B-}

\maketitle

\section{Introduction}

The study of interacting many-body systems remains one of the most active research fields in physics~\cite{Ringbook,Blaizotbook,Negelebook,Rammerbook}. The main computational challenge in electronic structure calculations is that a basis of many-body states, spanning the Hilbert space, becomes exponentially large with system size~\cite{Bartlett}. In many cases, however, one is not interested in all the information contained in the full many-body wave-function (or density matrix). Experimental observations are typically related to expectation values of certain one and two-body operators. This allows for various approximation schemes, which focus on correlation functions to retain reduced information on the state of the system. This is the basis for the density-functional-theory (DFT) and its time-dependent extension (TDDFT)~\cite{Casida,Bertsch2000,Takimoto2007,Gross,Onida2002}. 

The many body hierarchy of reduced density matrices can be truncated systematically by making a cumulant expansion of the wavefunction. Extensive work had been devoted to reduced description in terms of the coupled one and two-body density matrices~\cite{Mazziotti-book,Coleman}. The main difficulty in the direct computation of reduced density matrices has been the N-representability problem, namely the lack of exact conditions that guarantee that a given reduced density matrix can be obtained from an $N$ electron wavefunction.

In the TDHF method~\cite{Ringbook, Chernyak1996,Bertsch2000, Takimoto2007,Thouless,Blaizotbook,Tretiak2002,Mukamel1997} the state of the system is assumed to be given by a single Slater determinant, resulting in a closed system of equations for the occupied single-particle orbitals. 
This is equivalent to truncation at the level of the reduced single-particle density matrix~\cite{Thouless}.
TDHF is commonly used as a simple, affordable, approximation for the electronic excitations and the optical responce of a system~\cite{Tretiak2002}.
TDDFT response functions have the same formal structure and simplicity as TDHF ones, except that the burden
of the many body problem is shifted into the construction of the functional~\cite{Berman2003,Tretiak2003}.

The TDHF method can be recasted as a system of equations for a set of coordinates, $\rho_{\alpha \beta}=\left<\hat{c}_\alpha^\dagger \hat{c}_\beta \right>$~\cite{Chernyak1996,Tretiak1998},
which describe the reduced, single particle density matrix. These can be used to calculate  expectation values of any single body operator, such as the optical polarization.  Note that if the orbital space contain $n$
occupied and $m$ unoccupied orbitals, only $n m$ out of the $(n+m)^2$ elements of this reduced density matrix, namely the electron-hole excitations, are needed to represent the full density matrix, and calculate the response~\cite{Thouless}. 
The TDHF equations of motion are approximate, and systematic extentions are not obvious due to the absence of a simple small parameter.

Recently, an exact phase space representation for the many-body fermionic density matrix was developed by Corney and Drummond~\cite{Corney2004,Corney2006a,Corney2006b}. This method is similar in spirit to the coherent states in Hilbert space approach of Cahill and Glauber~\cite{Cahill1999}, since both employ an overcomplete basis set. In Corney and Drummond's approach the many body density matrix is expanded as an ensemble of Gaussian operators $\hat{\Lambda} ({\mathbf n})$  [Eq. (\ref{deflambda})] in the many-electron phase space,
\begin{equation}
\label{expand}
 \hat{\rho} = \int d {\mathbf n} P ({\mathbf n},t ) \hat{\Lambda} (\mathbf{n}),
\end{equation}
where $P ({\mathbf n},t )$ is a classical probability distribution for the matrix elements $n_{\alpha \beta}$. This density matrix is assumed to be block diagonal in Fock space, so that coherences are only allowed between states with the same number of particles, which is adequate for most applications. Coherences between states differing by an even number of particles, resulting in anomalous correlations, can be included as well.~\cite{Corney2004,Corney2006a,Corney2006b} Grassmann variables, which are essential for the coherent state representation in Hilbert space, are avoided, thus providing a more intuitive physical interpretation of various quantities. The exact time evolution of the probability distribution $P({\mathbf n},t)$ for fermions with two body interactions is described by a Fokker-Planck equation. The dynamics of the many-body system is thus mapped onto an ensemble of stochastic trajectories $n_{\alpha \beta} (t)$ in real (or imaginary) time. Gaussian phase space operator representation have been also used to derive the Hartree Fock Bogoliubov equations for Bose Einstein condensates ~\cite{Chernyak2003}. 

So far the Gaussian phase-space representation (GPSR) for fermions has mostly been used to study the ground states of  Hubbard-like models~\cite{Assaad2005, Aimi2007,Corboz2008}. Exceptions are the study of a mixed boson-fermion model of molecular dissociation~\cite{Corney2004} and a non interacting system~\cite{Corney2006b}. The GPSR can be used to study the time evolution of a system simply by using the Fokker-Planck equation to propagate $P({\mathbf n},t)$ in time. Therefore, this method maps the many body dynamics onto an evolution in a classical parameter space. The goal of the current paper is to show that the GPSR can be used to derive a new approximation scheme, which naturally connect to, and go beyond the TDHF. The possibility of using GSPR to solve the sign problem in imaginary time calculations is under debate.~\cite{Rombouts2006} However, in the real time dynamics considered here complex phases are unavoidable.

The Gaussian parameters $n_{\alpha \beta}$ play a role similar to the reduced single electron density matrix elements $\rho_{\alpha \beta}$ in TDHF. $P({\mathbf n})$ can be viewed as a distribution of  the matrix elements of the reduced single electron density matrix. We show that there exist a form of the stochastic equations for $n_{\alpha \beta} (t)$ whose deterministic part, the non random drift terms, coincides with the TDHF equations. It should be noted that the Gaussian operator basis is overcomplete, and thus allows for many equivalent forms of the stochastic equations. This freedom may be used to simplify the implementation of the method. 

A practical, exact, numerical scheme for computing the stochastic trajectories in real time is yet to be developed. 
We use the GPSR to construct new types of approximations for the dynamics of excitations.
By assuming that the probability distribution of the parameters $n_{\alpha \beta}$ is Gaussian, we obtain a closed system of equations for the single-particle and two-particle reduced density matrices, which is a direct extension of TDHF.
All approaches for performing many body computations by using single and two particles density matrices suffer from the $N$ representability problem, i.e., it is not guaranteed that the approximate reduced density matrices can be derived from an $N$ electron wavefunction~\cite{Mazziotti-book}. 
The GPSR of the density matrix in Fock space implies that $N$ has a distribution. Developing constraints that restrict Eq. (\ref{expand}) to a pure state described by a wavefunction with $N$ electrons will be of interest for molecular applications.
This limitation is less severe for large systems where  a distribution of $N$ makes a minimal effect. or for open systems like molecules in junctions.

In Sec.~\ref{tdhfsec} we present the TDHF equations of motion of the reduced density matrix. This sets the stage for the other methods. In Sec.~\ref{stochastic} we present the GPSR of Corney and Drummond~\cite{Corney2004,Corney2006a,Corney2006b} and derive the Fokker-Planck equation. In Sec~\ref{number} we derive an expression for the probability distribution of the number of electrons, and examine the conditions which ensure that a representation correspond to a state with a given particle number.
In Sec.~\ref{hierarchy} we develop an approximate truncated hierarchy, which goes beyond TDHF, by assuming that the probability distribution $P({\bf n},t)$ in Eq. (\ref{expand}) has a Gaussian form.  In Sec. {\ref{unraveling}} we present stochastic equations of motion, which are equivalent to the Fokker-Planck equation, and compare them to the TDHF equation derived in Sec.~\ref{tdhfsec}.
Our results are summarized in Sec.~\ref{disc}.

\section{The time dependent Hartree-Fock approximation}
\label{tdhfsec}

We consider a many-fermion system with two-body interactions, whose Hamiltonian is given by
\begin{equation}
 \hat{\cal H} = \sum_{\alpha \beta} t_{\alpha \beta} \hat{c}^\dagger_\alpha \hat{c}_\beta + \sum_{\alpha \beta \gamma \delta} V_{\alpha \beta \gamma \delta} \hat{c}_\alpha^\dagger \hat{c}_\beta^\dagger \hat{c}_\gamma \hat{c}_\delta.
\label{universalh}
\end{equation}
The indices $\alpha,\beta,\gamma,\delta$ denote an orthogonal one particle basis of spin orbitals. The creation and annihilation operators satisfy the Fermi anti-commutation rule
\begin{equation}
 \hat{c}_\alpha^\dagger \hat{c}_\beta + \hat{c}_\beta \hat{c}_\alpha^\dagger = \delta_{\alpha \beta}.
\end{equation}
Without restricting the generality, the two-body interaction $V_{\alpha \beta \gamma \delta}$ is taken to be anti-symmetric with respect to permutation of the indices $\alpha$ and $\beta$ or $\gamma$ and $\delta$. 

Our goal is to derive an equation of motion for a reduced single particle density matrix
\begin{equation}
 \rho_{\alpha \beta} \equiv \left< \hat{c}^\dagger_\alpha \hat{c}_\beta \right>= \left< \Phi (t) \right| \hat{c}^\dagger_\alpha \hat{c}_\beta \left| \Phi (t) \right>,
\end{equation}
where $\left| \Phi (t) \right>$ is the many body wavefunction. We start from the Heisenberg equation
\begin{equation}
\label{tdhf1}
 \dot{\rho}_{\epsilon \zeta} = i \left< \Phi (t) \right| \left[ \hat{\cal H}, \hat{c}^\dagger_\epsilon \hat{c}_\zeta \right] \left| \Phi (t) \right>,
\end{equation}
where we work in units such that $\hbar=1$.

The commutator in Eq. (\ref{tdhf1}) is easily evaluated, leading to
\begin{eqnarray}
 \dot{\rho}_{\epsilon \zeta} & = & -i \sum_\gamma \left( t_{\zeta \gamma}\rho_{\epsilon \gamma}-t_{\gamma \epsilon} \rho_{\gamma \zeta} \right)  \nonumber
 -  i \sum_{\alpha \beta \gamma \delta} V_{\alpha \beta \gamma \delta} \left\{ \left<\Phi(t)\right| \hat{c}^\dagger_\alpha \hat{c}_{\beta}^\dagger \hat{c}_\delta \hat{c}_\zeta \left| \Phi (t) \right> \delta_{\epsilon \gamma} \right. \\ & - & \left. \left<\Phi(t)\right| \hat{c}^\dagger_\alpha \hat{c}_{\beta}^\dagger \hat{c}_\gamma \hat{c}_\zeta \left| \Phi (t) \right> \delta_{\epsilon \delta}+\left<\Phi(t)\right| \hat{c}^\dagger_\epsilon \hat{c}_{\beta}^\dagger \hat{c}_\gamma \hat{c}_\delta \left| \Phi (t) \right> \delta_{\zeta \alpha}-\left<\Phi(t)\right| \hat{c}^\dagger_\epsilon \hat{c}_{\alpha}^\dagger \hat{c}_\gamma \hat{c}_\delta \left| \Phi (t) \right> \delta_{\zeta \beta}\right\}.
\label{tdhf2}
\end{eqnarray}
The TDHF approximation assumes that the state of the system is given by a single slater determinant $\left| \Phi (t)\right>$, at all times. Using Wick's theorem, we replace the two-body matrix elements with products of single particle matrix elements,
\begin{equation}
 \left<\Phi(t)\right| \hat{c}^\dagger_\alpha \hat{c}_{\beta}^\dagger \hat{c}_\gamma \hat{c}_\delta \left| \Phi (t) \right> = \rho_{\alpha \delta} \rho_{\beta \gamma} - \rho_{\alpha \gamma} \rho_{\beta \delta}. \label{hfapprox}
\end{equation}
Substituting Eq. (\ref{hfapprox}) into Eq. ({\ref{tdhf2}}) results in the TDHF equations for the reduced density matrix,
\begin{eqnarray}
 \dot{\rho}_{\epsilon \zeta} & = & -i \sum_\gamma \left( t_{\zeta \gamma}\rho_{\epsilon \gamma}-t_{\gamma \epsilon} \rho_{\gamma \zeta} \right)  -i \sum_{\alpha \beta \gamma} \rho_{\alpha \beta} \rho_{\gamma \zeta} \left( V_{\alpha \gamma \beta \epsilon} - V_{\gamma \alpha \beta \epsilon} + V_{\gamma \alpha \epsilon \beta} - V_{\alpha \gamma \epsilon \beta} \right) \nonumber \\
& - & i \sum_{\alpha \beta \delta} \rho_{\alpha \beta} \rho_{\epsilon \delta} \left( V_{\alpha \zeta \delta \beta }-V_{\alpha \zeta \beta \delta}+V_{\zeta \alpha \beta \delta} - V_{\zeta \alpha \delta \beta}\right). \label{tdhf3}
\end{eqnarray}
Eq. (\ref{tdhf3}) can be further simplified using the antisymmetry of $V_{\alpha \beta \gamma \delta}$ to permutations of $\alpha$ and $\beta$, or $\gamma$ and $\delta$. This gives
\begin{equation}
 \dot{\rho}_{\epsilon \zeta}  =  -i \sum_\gamma \left( t_{\zeta \gamma}\rho_{\epsilon \gamma}-t_{\gamma \epsilon} \rho_{\gamma \zeta} \right) -4i \sum_{\alpha \beta \gamma} \rho_{\alpha \beta} \rho_{\gamma \zeta}  V_{\alpha \gamma \beta \epsilon} -4 i \sum_{\alpha \beta \delta} \rho_{\alpha \beta} \rho_{\epsilon \delta} V_{\alpha \zeta \delta \beta }. \label{tdhffinal}
\end{equation}

 Eq. (\ref{tdhffinal}) implies that $\rho_{\alpha \beta}$ can be viewed as classical oscillator coordinates, which follow a deterministic trajectory~\cite{Chernyak1996,Tretiak1998}.   The TDHF equations will be generalized in Sec.~\ref{hierarchy} using the phase-space representation of a fermionic system, which retains the same number of variables, $~N^2$, as the TDHF but treats them as stochastic coordinates. We must then work with their distribution rather than with deterministic trajectories in that space.

\section{Gaussian phase-space representation for fermions}
\label{stochastic}

 In this section we briefly present the main results of the GPSR, without proofs, following Refs.~\onlinecite{Corney2006a,Corney2006b}. 
We start by introducing the Gaussian operators 
\begin{equation}
\label{deflambda}
 \hat{\Lambda} (\mathbf n) \equiv \det \tilde{\mathbf n} : \exp \left( - \hat{\mathbf c}^\dagger \left[ 2 {\mathbf I} - \tilde{\mathbf n}^{-T} \right] \hat{\mathbf c} \right) :,
\end{equation}
where $\tilde{\mathbf n}= {\mathbf I}-{\mathbf n}$ is a square matrix of parameters whose size is given by the spin orbitals basis set, and $: \cdots :$ denotes normal ordering. A normal-ordered product of creation and annihilation operators is one where the creation operators are to the left of the annihilation operators. The sign must be changed when two operators are interchanged during reordering. (For instance, $: \hat{c}_\alpha \hat{c}_\beta^\dagger: = -\hat{c}_\beta^\dagger \hat{c}_\alpha. $ )
The operators $\hat{\Lambda}$ have the following useful properties
\begin{eqnarray}
 \trace \hat{\Lambda} & = & 1, \\
\trace \left( \hat{\Lambda} \hat{c}^\dagger_\alpha \hat{c}_\beta \right) & = & n_{\alpha \beta}, \label{getn} \\
\trace \left( \hat{\Lambda} \hat{c}^\dagger_\alpha \hat{c}_\beta^\dagger \hat{c}_\gamma \hat{c}_\delta \right) & = & n_{\alpha \delta} n_{\beta \gamma}-n_{\alpha \gamma} n_{\beta \delta}, \label{wick2}\\
\hat{c}^\dagger_\alpha \hat{c}_\beta \hat{\Lambda} & = & n_{\alpha \beta} \hat{\Lambda} + \sum_{\gamma \delta} \tilde{n}_{\alpha \gamma} n_{\delta \beta} \frac{\partial \hat{\Lambda}}{\partial n_{\delta \gamma}}, \label{ident1} \\
\hat{\Lambda} \hat{c}^\dagger_\alpha \hat{c}_\beta & = & n_{\alpha \beta} \hat{\Lambda} + \sum_{\gamma \delta} n_{\alpha \gamma} \tilde{n}_{\delta \beta} \frac{\partial \hat{\Lambda}}{\partial n_{\delta \gamma}},  \label{ident2}
\end{eqnarray}
where $\trace$ denotes the trace is in the many-body Hilbert space.
Generally, the Gaussian operators also obey Wick's theorem,
\begin{equation}
 \label{wicks}
\trace \left[ : \hat{a}_{\mu_1} \cdots \hat{a}_{\mu_{2r}} \hat{\Lambda}: \right] = \sum_{P} (-1)^P \trace \left[ : \hat{a}_{\nu_1}  \hat{a}_{\nu_{2}} \hat{\Lambda}: \right] \times \cdots \times \trace \left[ : \hat{a}_{\nu_{2r-1}} \hat{a}_{\nu_{2r}} \hat{\Lambda}: \right],
\end{equation}
where $\hat{\bf a}$ is a vector composed of all creation and annihilation operators, and $\nu_j=\mu_{P(j)}$. The sum runs over all the distinct pair permutations. It should be noted that the normal ordering convention used in Eq. (\ref{wicks}) is that only the operators which do not belong to $\hat{\Lambda}$ are ordered so that all the creation operators are to its left, and the annihilation operators to its right. Eq. (\ref{wick2}) is a special case of Eq. (\ref{wicks}).
Proofs, as well as several other similar relations, can be found in Ref.~\onlinecite{Corney2006a}. Equations (\ref{ident1}) and (\ref{ident2}) allow to replace the action of creation and annihilation operators on $\hat{\Lambda}$ by derivatives with respect to the parameters $n_{\alpha \beta}$. 

With the help of Eqs. (\ref{expand}) and (\ref{getn}) we see that the reduced single particle density matrix is given by the first moment of the distribution $P({\mathbf n})$,
\begin{equation}
 \rho_{\alpha \beta} = \int n_{\alpha \beta} \; P({\mathbf n},t)  d {\mathbf n} = \left< n_{\alpha \beta} \right>_P. 
\label{rhovian}
\end{equation}
We introduce the following notation, for a fermion operator $\hat{A}$
\begin{equation}
 \left< \hat{A} \right> \equiv \int d {\mathbf n} P({\mathbf n},t) \trace \left[ \hat{A} \hat{\Lambda} ({\mathbf n}) \right] \equiv \left< \trace \hat{A} \hat{\Lambda} ({\mathbf n}) \right>_P.
\end{equation}
Note that there is a double averaging here. The $\trace$ takes care of the quantum average
for a given set of parameters ${\mathbf n}$. We then apply a classical average  $\left< \dots \right>_P$ over the phase space distribution $P({\mathbf n})$.

The reduced $p$-particle density matrix depends on the lowest $p$ moments of $P({\mathbf n},t)$. We will use the expressions for expectation values of two and three body operators
\begin{eqnarray}
{\cal M}_{\alpha \beta, \gamma \delta} & \equiv & \left< \hat{c}^\dagger_\alpha \hat{c}^\dagger_\beta \hat{c}_\gamma \hat{c}_\delta \right>=  \left< n_{\beta \gamma} n_{\alpha \delta} - n_{\alpha \gamma } n_{\beta \delta}\right>_P, \label{avgp2} \\
{\cal Y}_{\alpha \beta \gamma, \delta \epsilon \zeta} & \equiv & \left< \hat{c}^\dagger_\alpha \hat{c}^\dagger_\beta \hat{c}^\dagger_\gamma \hat{c}_\delta \hat{c}_\epsilon \hat{c}_\zeta \right> \nonumber \\ & = & \left< n_{\alpha \zeta} n_{\beta \epsilon} n_{\gamma \delta} - n_{\alpha \delta} n_{\beta \epsilon} n_{\gamma \zeta} + n_{\alpha \delta} n_{\beta \zeta} n_{\gamma \epsilon} -n_{\alpha \epsilon} n_{\beta \zeta} n_{\gamma \delta}+n_{\alpha \epsilon} n_{\gamma \zeta} n_{\beta \delta}- n_{\alpha \zeta} n_{\beta \delta} n_{\gamma \epsilon}\right>_P. \label{avgp3}
\end{eqnarray}

Before turning to derive the equation of motion for $P({\mathbf n},t)$, it is important to note that the representation Eq.~(\ref{expand}) is not unique. More than one probability distribution $P({\mathbf n},t)$ may represent the same many-particle density matrix, due to the overcompleteness of the phase space representation. This is easily demonstrated by the fact that the many body density matrix of a system with $M$ orbitals only depends on the lowest $M$'th moments of $P({\mathbf n})$. The higher moments can thus be arbitrarily chosen. In the following we present the simplest derivation of the Fokker-Planck equation, which is naturally related to the TDHF. Alternative Fokker-Planck equations, whose solutions constitute different representations of the same many body density matrix, are discussed in App.~\ref{gauge}.

Substitution of Eq. (\ref{expand}) into the Liouville equation 
\begin{equation}
 \frac{\partial \hat{\rho}}{\partial t} = -i \left[ \hat{\cal H}, \hat{\rho}\right],
\end{equation}
gives
\begin{equation}
 \int \frac{\partial P({\mathbf n},t)}{\partial t} \hat{\Lambda}({\mathbf n}) d {\mathbf n} = -i \int P({\mathbf n},t) \left[\hat{\cal H},\hat{\Lambda}({\mathbf n})\right] d {\mathbf n}.
\label{tempeq}
\end{equation}
The identities (\ref{ident1}) and (\ref{ident2}) can now be used to replace the commutator in Eq. (\ref{tempeq}) with a differential operator. 
To that end, we decompose the matrix $V_{\alpha \beta \gamma \delta}$ in the form
\begin{equation}
\label{defQ}
 V_{\alpha \beta \gamma \delta} = \frac{i}{2} \sum_c Q_{\alpha \gamma,c} Q_{\beta \delta,c}.
\end{equation}
 It should be always possible to find such a decomposition. This decomposition and the number of components $c$ are not unique, and the latter can be chosen to be arbitrarily large. This decomposition eventually leads to a Fokker-Planck equation whose diffusion matrix is manifestly positive definite. 
For Hamiltonians with two-body interactions Eqs (\ref{ident1}) and (\ref{ident2}) this leads to a differential operator with second order derivatives
\begin{equation}
 -i \left[\hat{\cal H}, \hat{\Lambda} \right] = \sum_{\alpha \beta} {\cal A}_{\alpha \beta} \frac{\partial \hat{\Lambda}}{\partial n_{\alpha \beta}}+\frac{1}{2} \sum_{\alpha \beta \gamma \delta} \sum_{c} {\cal B}^{(1)}_{\alpha \beta,c} {\cal B}^{(1)}_{\gamma \delta,c} \frac{\partial^2 \hat{\Lambda}}{\partial n_{\alpha \beta} \partial n_{\gamma \delta}}+\frac{1}{2} \sum_{\alpha \beta \gamma \delta} \sum_{c} {\cal B}^{(2)}_{\alpha \beta,c} {\cal B}^{(2)}_{\gamma \delta,c} \frac{\partial^2 \hat{\Lambda}}{\partial n_{\alpha \beta} \partial n_{\gamma \delta}}.
\label{commute}
\end{equation}

Closed expressions for ${\cal A}_{\alpha \beta}$ and ${\cal B}_{\alpha \beta}^{(1,2)}$, starting with
the Hamiltonian Eq. (\ref{universalh}), are obtained by a straightforward but tedious calculation. 
The decomposition (\ref{defQ}) allows to write the terms with second order derivatives in the form of Eq. (\ref{commute}), with
\begin{eqnarray}
 {\cal B}^{(1)}_{\epsilon \zeta,c} & = & \sum_{\alpha \beta} Q_{\alpha \beta,c} n_{\alpha \zeta} \tilde{n}_{\epsilon \beta} \\
  {\cal B}^{(2)}_{\epsilon \zeta,c} & = &i \sum_{\alpha \beta} Q_{\alpha \beta,c} \tilde{n}_{\alpha \zeta} n_{\epsilon \beta},
\end{eqnarray}
where $\tilde{n}$ was defined after Eq. (\ref{deflambda}).
Furthermore, a lengthy calculation results in
\begin{eqnarray}
  {\cal A}_{\epsilon \zeta} & = & -i \sum_\gamma \left( t_{\zeta \gamma} n_{\epsilon \gamma}-t_{\gamma \epsilon} n_{\gamma \zeta} \right)  -i \sum_{\alpha \beta \gamma} n_{\alpha \beta} n_{\gamma \zeta} \left( V_{\alpha \gamma \beta \epsilon} - V_{\gamma \alpha \beta \epsilon} + V_{\gamma \alpha \epsilon \beta} - V_{\alpha \gamma \epsilon \beta} \right) \nonumber \\
& - & i \sum_{\alpha \beta \delta} n_{\alpha \beta} n_{\epsilon \delta} \left( V_{\alpha \zeta \delta \beta }-V_{\alpha \zeta \beta \delta}+V_{\zeta \alpha \beta \delta} - V_{\zeta \alpha \delta \beta}\right). 
\label{Aito}
\end{eqnarray}

By substituting (\ref{commute}) in Eq. (\ref{tempeq}), and integrating by parts, one obtains a Fokker-Planck-type equation for $P({\mathbf n},t)$. However, one should note that the coefficients ${\cal A}_{\alpha \beta}$ and ${\cal B}^{(1,2)}_{\alpha \beta}$ are complex valued rather than real, and it is not clear at this point that the term with the second order derivatives is positive definite. A representation with real parameters can be obtained by using the analyticity of $\hat{\Lambda} (\mathbf n)$, and treating the real and the imaginary parts of $n_{\alpha \beta}=n_{\alpha,\beta}^{(x)}+i n_{\alpha \beta}^{(y)}$ as independent variables. The identity $\partial \hat{\Lambda}/\partial n_{\alpha \beta}=\partial \hat{\Lambda}/\partial n_{\alpha \beta}^{(x)}=-i\partial \hat{\Lambda}/\partial n_{\alpha \beta}^{(y)}$ can then be used to write
\begin{equation}
 {\cal A}_{\alpha \beta} \frac{\partial \hat{\Lambda}}{\partial n_{\alpha \beta}} = {\cal A}_{\alpha \beta}^{(x)} \frac{\partial \hat{\Lambda}}{\partial n_{\alpha \beta}^{(x)}}+{\cal A}_{\alpha \beta}^{(y)} \frac{\partial \hat{\Lambda}}{\partial n_{\alpha \beta}^{(y)}}.
\end{equation}
Here ${\cal A}^{(x)}$ (${\cal A}^{(y)}$) denotes the real (imaginary) parts of ${\cal A}$ respectively.
A similar relation holds for the terms involving second derivatives and for ${\cal B}^{(1,2)}$.

After separating the real and imaginary parts of ${\cal A}$ and ${\cal B}$, and performing an integration by parts, one obtains
\begin{eqnarray}
\label{FPeq}
 \frac{\partial P}{\partial t} & = & -\sum_{\alpha \beta}\left( \frac{\partial}{\partial n_{\alpha \beta}^{(x)}} {\cal A}_{\alpha \beta}^{(x)} P+\frac{\partial}{\partial n_{\alpha \beta}^{(y)}} {\cal A}_{\alpha \beta}^{(y)} P  \right) \\
& + & \frac{1}{2} \sum_{\alpha \beta \gamma \delta} \sum_c \left( \frac{\partial^2}{\partial n_{\alpha \beta}^{(x)}\partial n_{\gamma \delta}^{(x)}} {\cal B}_{\alpha \beta ,c}^{(1,x)} {\cal B}_{\gamma \delta ,c}^{(1,x)} P + \frac{\partial^2}{\partial n_{\alpha \beta}^{(y)}\partial n_{\gamma \delta}^{(y)}} {\cal B}_{\alpha \beta ,c}^{(1,y)} {\cal B}_{\gamma \delta ,c}^{(1,y)} P+2 \frac{\partial^2}{\partial n_{\alpha \beta}^{(x)}\partial n_{\gamma \delta}^{(y)}} {\cal B}_{\alpha \beta ,c}^{(1,x)} {\cal B}_{\gamma \delta ,c}^{(1,y)} P\right) \nonumber \\
& + & \frac{1}{2} \sum_{\alpha \beta \gamma \delta} \sum_c \left( \frac{\partial^2}{\partial n_{\alpha \beta}^{(x)}\partial n_{\gamma \delta}^{(x)}} {\cal B}_{\alpha \beta ,c}^{(2,x)} {\cal B}_{\gamma \delta ,c}^{(2,x)} P + \frac{\partial^2}{\partial n_{\alpha \beta}^{(y)}\partial n_{\gamma \delta}^{(y)}} {\cal B}_{\alpha \beta ,c}^{(2,y)} {\cal B}_{\gamma \delta ,c}^{(2,y)} P+2 \frac{\partial^2}{\partial n_{\alpha \beta}^{(x)}\partial n_{\gamma \delta}^{(y)}} {\cal B}_{\alpha \beta ,c}^{(2,x)} {\cal B}_{\gamma \delta ,c}^{(2,y)} P\right) \nonumber.
\end{eqnarray}
Eq. (\ref{FPeq}) is a positive definite Fokker-Planck equation, and the probability distribution $P({\mathbf n},t)$ is guaranteed to remain non-negative at all times. In the derivation of Eq. (\ref{FPeq}) boundary terms in the integration by parts were neglected. When boundary terms are finite, the methods discussed in App.~\ref{gauge} can be used to derive equivalent Fokker-Planck equations with vanishing boundary terms.

\section{Statistical properties of the number of particles in the Gaussian phase-space representation}
\label{number}

The Hamiltonian (\ref{universalh}) conserves the number of electrons. However the distribution function $P({\mathbf n})$ may represent a statistical mixture of states with different numbers of electrons. It is of interest to find conditions which ensure that $P({\mathbf n})$ represent a fixed number a system with $N$ electrons. So far the GPSR was used to study systems where knowledge about the distribution of the number of electrons was not crucial. 
In this section we derive conditions which guarantee that $P({\mathbf n})$ represents a state with a given number of electrons. This will be needed for application to isolated molecules. Furthermore, for general states represented by $P({\mathbf n})$,
we derive expressions for the probability to find $k$ electrons in the system, $\tilde{P} (k)$.

We start by calculating the distribution of values of $\hat{N}=\sum_\alpha \hat{c}^\dagger_\alpha \hat{c}_\alpha$, the number operator,
\begin{equation}
\label{deffn}
 f(N) \equiv \left< \delta \left( N - \hat{N} \right) \right>.
\end{equation}
It will be convenient to compute the generating function
\begin{equation}
\label{defgs}
 G(s) = \int d N e^{- i s N} f(N) = \left< e^{- i s \hat{N}} \right> .
\end{equation}
With the help of Eq. (\ref{ident1}), it is easy to show that
\begin{equation}
 \hat{N} \hat{\Lambda} = \left( \sum_{\alpha} n_{\alpha \alpha} + \sum_{\alpha \mu \nu} \tilde{n}_{\alpha \mu} n_{\nu \alpha} \frac{\partial}{\partial n_{\nu \mu}} \right) \hat{\Lambda}.
\end{equation}
As a result
\begin{equation}
 G(s) = \trace \int d {\mathbf n} P ({\mathbf n}) \exp \left[-is \left( \sum_{\alpha} n_{\alpha \alpha} + \sum_{\alpha \mu \nu} \tilde{n}_{\alpha \mu} n_{\nu \alpha} \frac{\partial}{\partial n_{\nu \mu}} \right) \right] \hat{\Lambda}.
\end{equation}
By using the Baker-Campbell-Hausdorff formula we then obtain
\begin{equation}
 G(s)=\left< \exp \left[ - i s A({\mathbf n}) - \frac{1}{2} s^2 B({\mathbf n}) \right] \right>_P
\end{equation}
where $A({\mathbf n}) = \smalltrace ({\mathbf n})$ and $B({\mathbf n}) = \smalltrace ({\mathbf n})-\smalltrace ({\mathbf n}^2)$, where $\smalltrace$ denotes the trace in the single particle Hilbert space.
(Recall that a trace in the many body space is denoted by $\trace$.)

Assuming that the integrations over ${\mathbf n}$ converge, we obtain the probability distribution (\ref{deffn}) by the inverse Fourier transform of Eq. (\ref{defgs})
\begin{equation}
 f(N) = \frac{1}{\sqrt{2 \pi}} \left< \frac{1}{\sqrt{B({\mathbf n})}} \exp \left[- \frac{(N-A({\mathbf n}))^2}{2 B({\mathbf n})} \right]\right>_P.
\end{equation}
The distribution $P({\mathbf n})$ will represent a state with a given number of electrons,  $N_0$, when $f(N)=\delta (N-N_0)$. This requires that  $P({\mathbf n})>0$ only for points ${\mathbf n}$ which satisfy
\begin{equation}
 \smalltrace ({\mathbf n}) = \smalltrace ({\mathbf n}^2) = N_0.
\end{equation}

Distributions $P({\mathbf n})$ which do not satisfy this conditions represent statistical mixtures of different number of electrons. To calculate the probability to find $k$ electrons, $\tilde{P} (k)$, we note that the Gaussian operators have a block structure in Fock space, with coherences only between states with the same number of particles. For instance, the Gaussian operator of a system with two orbitals is~\cite{Corney2006a}
\begin{eqnarray}
 \hat{\Lambda}({\mathbf n}) &  = & \det \tilde{\mathbf n} \left| 00 \right> \left< 00 \right| + \left(n_{11} \tilde{n}_{22}+n_{12} n_{21} \right) \left| 10 \right> \left< 10 \right| + \left( \tilde{n}_{11} n_{22}+n_{12} n_{21}\right) \left| 01 \right> \left< 01 \right| + \det {\mathbf n} \left| 11 \right> \left< 11 \right| \\ &+ &n_{21} \left| 10 \right> \left< 01 \right| + n_{12} \left| 01 \right> \left< 10 \right|. \nonumber
\end{eqnarray}
The probability to find the system with any number of particles can be found by taking the coefficients of the corresponding populations and averaging over $P({\mathbf n})$. This can be done by defining
\begin{equation}
 {\cal C}_l \equiv \trace \hat{\rho} \sum_{\alpha_1, \cdots, \alpha_l} \hat{c}^\dagger_{\alpha_1} \cdots \hat{c}^\dagger_{\alpha_l} \hat{c}_{\alpha_l} \cdots \hat{c}_{\alpha_1},
\end{equation}
for $l=1,2,\cdots,M$, where $M$ denotes the basis size (i.e. the number of orbitals). Only populations of states with at least $l$ electrons contribute to ${\cal C}_l$. A direct calculation gives
\begin{equation}
 {\cal C}_l = \sum_{k=l}^M \frac{k!}{(k-l)!} \tilde{P} (k).
\end{equation}
This relation can be inverted, leading to
\begin{equation}
\label{pp1}
\tilde{P} (k) = \frac{1}{k!} \sum_{l=k}^M (-1)^{l-k} \frac{1}{(l-k)!} {\cal C}_l,
\end{equation}
which is valid for $k=1,2,\cdots,M$.

Wick's theorem (\ref{wicks}) can be used to write
\begin{equation}
 \label{pp2}
{\cal C}_l = l!  \sum_{(\alpha_1,\cdots,\alpha_l)}\left< \det n \left[ \alpha_1,\cdots,\alpha_l \right] \right>_P,
\end{equation}
where sum is over all {\em ordered} sets of indices $(\alpha_1,\cdots,\alpha_l)$, while ${\mathbf n} \left[ \alpha_1, \cdots, \alpha_l \right]$ is the minor of ${\mathbf n}$ obtained by deleting all rows and columns except the ones whose indices are $\alpha_1, \cdots, \alpha_l$.

Equations (\ref{pp1}) and (\ref{pp2}) combine to give
the probability to find $k$ electrons in the system 
\begin{equation}
\label{getpn}
 \tilde{P} (k) = \sum_{l=k}^M (-1)^{l-k} \left( \begin{array}{c}
 ×l \\ k 
\end{array}
\right)
\sum_{(\alpha_1,\cdots,\alpha_l)} \left< \det {\mathbf n} \left[ \alpha_1, \cdots, \alpha_l \right]\right>_P,
\end{equation}
which is valid for $k=1,2, \cdots , M$.  Equation (\ref{getpn}) is especially simple for the probability that all the orbitals are filled
\begin{equation}
 \tilde{P} (M) = \left< \det {\mathbf n} \right>_P.
\end{equation}
We obtain the probability that there are no electrons (all orbitals are unoccupied) by using the normalization condition
\begin{equation}
 \tilde{P} (0)=1-\sum_{k=1}^M \tilde{P} (k) = 1 + \sum_{l=1}^M (-1)^l \sum_{(\alpha_1,\cdots,\alpha_l)} \left< \det {\mathbf n} \left[ \alpha_1, \cdots, \alpha_l \right]\right>_P = \left< \det \tilde{\mathbf n}\right>_P.
\end{equation}

In this section we have investigated the correspondence between the phase space distribution $P({\bf n})$ and the number of particles in the system. This is one aspect of a broader problem, namely, how to constrain $P({\bf n})$ and its phase space evolution, so that it will correspond to a density matrix which exhibits a certain physical property. For instance, what are the conditions on $P({\bf n})$ so that it corresponds to a pure state, or to a single slater determinant? The identification of such constraints is an open question.

\section{Truncating the hierarchy for one and two-body density matrices}
\label{hierarchy}

In this section we use the exact GPSR of the many-body dynamics to develop new approximation schemes which provide a natural extension of the TDHF approximation. 

The time evolution of expectation values of any operator follows the Heisenberg equation
\begin{equation}
\label{traceo}
 \frac{d}{dt} \trace \hat{\rho} \hat{\cal O} = i \trace \hat{\rho} \left[\hat{\cal H},\hat{\cal O} \right]. 
\end{equation}
For Hamiltonians of the form (\ref{universalh}), substitution of normally ordered products of equal numbers of creation and annihilation operators in Eq. (\ref{traceo}) results in the usual many body hierarchy of equations of motion. The $N$'th equation in the hierarchy gives the time derivative of an $N$-particle reduced density matrix in terms the reduced density matrices of up to $N+1$ particles. 
The first two members in this hierarchy are given by
\begin{equation}
\label{hirarchy1}
 \frac{d}{dt} \rho_{\epsilon \zeta}  =  i \left\{ \sum_{\alpha \beta} t_{\alpha \beta} \left( \rho_{\alpha \zeta} \delta_{\beta \epsilon} - \rho_{\epsilon \beta} \delta_{\zeta \alpha} \right)
  +  \sum_{\alpha \beta \gamma \delta} V_{\alpha \beta \gamma \delta} \left( {\cal M}_{\alpha \beta , \gamma \zeta} \delta_{\delta \epsilon} -{\cal M}_{\alpha \beta , \delta \zeta} \delta_{\gamma \epsilon} + {\cal M}_{\epsilon \alpha , \gamma \delta} \delta_{\zeta \beta}-{\cal M}_{\epsilon \beta , \gamma \delta}\delta_{\zeta \alpha}\right) \right\} 
\end{equation}
and 
\begin{eqnarray}
\label{hirarchy2}
 \frac{d}{dt} {\cal M}_{\epsilon \zeta , \mu \nu} & = & i \sum_{\alpha \beta} t_{\alpha \beta} \left[{\cal M}_{\alpha \zeta , \mu \nu} \delta_{\beta \epsilon} - {\cal M}_{\alpha \epsilon, \mu \nu} \delta_{\beta \zeta} +{\cal M}_{\epsilon \zeta , \nu \beta} \delta_{\mu \alpha} - {\cal M}_{\epsilon \zeta , \mu \beta} \delta_{\nu \alpha}\right] \\
& + & i \sum_{\alpha \beta \gamma \delta} V_{\alpha \beta \gamma \delta} \left[ {\cal Y}_{\alpha \beta \epsilon, \gamma \mu \nu} \delta_{\delta \zeta}-{\cal Y}_{\alpha \beta \epsilon, \delta \mu \nu}\delta_{\gamma \zeta} + {\cal Y}_{\alpha \beta \zeta , \delta \mu \nu} \delta_{\gamma \epsilon} -{\cal Y}_{\alpha \beta \zeta , \gamma \mu \nu} \delta_{\delta \epsilon} 
+  {\cal Y}_{\epsilon \zeta \alpha , \nu \gamma \delta} \delta_{\mu \beta} \right. \nonumber \\  & - & \left. {\cal Y}_{\epsilon \zeta \alpha , \mu \gamma \delta} \delta_{\nu \beta} +{\cal Y}_{\epsilon \zeta \beta , \mu \gamma \delta} \delta_{\nu \alpha}- {\cal Y}_{\epsilon \zeta \beta , \nu \gamma \delta} \delta_{\mu \alpha} 
+  {\cal M}_{\alpha \beta , \mu \nu} \left( \delta_{\delta \epsilon} \delta_{\gamma \zeta} - \delta_{\gamma \epsilon} \delta_{\delta \zeta} \right) +{\cal M}_{\epsilon \zeta , \gamma \delta} \left( \delta_{\mu \alpha} \delta_{\nu \beta} - \delta_{\nu \alpha} \delta_{\mu \beta} \right) \right], \nonumber
\end{eqnarray}
where $\rho$, ${\cal M}$ and $\cal Y$ were defined in Eqs. (\ref{rhovian}), (\ref{avgp2}) and (\ref{avgp3}).

The hierarchy for the reduced single-particle and two-particle density matrices will be closed by assuming that the probability distribution $P({\mathbf n})$ is Gaussian. 
 Such an approximation is similar in spirit to the Hartree-Fock approximation, since it makes an ansatz on the time dependent state of the system and it includes a single Slater determinant as a special case.
However, it is more general, since it takes some two body correlations into account. Due to the lack of a variational principle, it is not guaranteed that the Gaussian approximation in phase space improves upon the Hartree-Fock approximation. However, it is likely to do so.

When $P({\mathbf n})$ has a Gaussian form, averages which are cubic in $n_{\alpha \beta}$, such as the ones appearing in Eq. (\ref{avgp3}), can be expressed in terms of the first and second moments of the distribution. For instance
\begin{equation}
\label{avgn3}
 \left< n_{\alpha \zeta} n_{\beta \epsilon} n_{\gamma \delta} \right>_P = \left< n_{\alpha \zeta} \right>_P \left< n_{\beta \epsilon} n_{\gamma \delta} \right>_P + \left<n_{\beta \epsilon} \right>_P \left< n_{\alpha \zeta} n_{\gamma \delta} \right>_P + \left<n_{\gamma \delta} \right>_P \left< n_{\alpha \zeta} n_{\beta \epsilon}\right>_P - 2 \left< n_{\alpha \zeta} \right>_P  \left<n_{\beta \epsilon} \right>_P \left<n_{\gamma \delta} \right>_P.
\end{equation}
 Substitution of (\ref{avgn3}) in Eq. (\ref{avgp3}), with the help of  Eqs. (\ref{avgp2}) and (\ref{rhovian}), leads to
\begin{eqnarray}
\label{c3element}
 {\cal Y}_{\alpha \beta \gamma , \delta \epsilon \zeta} & = & \rho_{\beta \epsilon} {\cal M}_{\alpha \gamma, \delta \zeta} + \rho_{\alpha \delta} {\cal M}_{\beta \gamma, \epsilon \zeta}+\rho_{\beta \zeta} {\cal M}_{\alpha \gamma,\epsilon \delta} + \rho_{\gamma \delta} {\cal M}_{\alpha \beta, \epsilon \zeta} + \rho_{\alpha \epsilon} {\cal M}_{\beta \gamma, \zeta \delta}+\rho_{\gamma \zeta} {\cal M}_{\alpha \beta , \delta \epsilon} \\ & + & \rho_{\beta \delta} {\cal M}_{\alpha \gamma , \zeta \epsilon}+ \rho_{\alpha \zeta} {\cal M}_{\beta \gamma , \delta \epsilon} + \rho_{\gamma \epsilon} {\cal M}_{\alpha \beta, \zeta \delta} -2 \left\{ \rho_{\alpha \zeta} \rho_{\beta \epsilon} \rho_{\gamma \delta} - \rho_{\alpha \delta} \rho_{\beta \epsilon} \rho_{\gamma \zeta} + \rho_{\alpha \delta} \rho_{\beta \zeta} \rho_{\gamma \epsilon} \right. \nonumber \\ & - & \left. \rho_{\alpha \epsilon} \rho_{\beta \zeta} \rho_{\gamma \delta} + \rho_{\alpha \epsilon} \rho_{\gamma \zeta} \rho_{\beta \delta} - \rho_{\alpha \zeta} \rho_{\beta \delta} \rho_{\gamma \epsilon} \right\}. \nonumber
\end{eqnarray}

Equation (\ref{c3element}) represent one out of many possible decoupling schemes which can be used to truncate the hierarchy of $N$-body reduced density matrices. It was derived by making an assumption on the form of $P({\bf n})$. Since it is based on an approximate ansatz on the many body density matrix it has an important advantage over other decoupling schemes: it ensures that the approximate density matrix is a physically allowed one, and therefore guarantees that the decoupling (\ref{c3element}) will not result in unphysical expectation values of operators. 

We can thus truncate the hierarchy at the $\rho$ and ${\cal M}$ level. Before doing that, we define the deviation of the two body correlation functions from its TDHF values
\begin{equation}
 \Delta {\cal M}_{\alpha \beta , \gamma \delta} \equiv {\cal M}_{\alpha \beta , \gamma \delta} -\rho_{\alpha \delta} \rho_{\beta \gamma} + \rho_{\alpha \gamma} \rho_{\beta \delta}.
\end{equation}
 By substitution of Eqs. (\ref{rhovian}), (\ref{avgp2}) and (\ref{c3element}) in Eqs. (\ref{hirarchy1}) and (\ref{hirarchy2}) we obtain

\begin{equation}
\label{eqrho}
 \dot{\rho}_{\epsilon \zeta} = -i \sum_\gamma \left( t_{\zeta \gamma} \rho_{\epsilon \gamma}-t_{\gamma \epsilon} \rho_{\gamma \zeta}\right)-4 i \sum_{\alpha \beta \gamma} V_{\alpha \gamma \beta \epsilon} \rho_{\alpha \beta} \rho_{\gamma \zeta} -4 i \sum_{\alpha \beta \delta} V_{\alpha \zeta \delta \beta} \rho_{\alpha \beta} \rho_{\epsilon \delta}-2i \sum_{\alpha \beta \delta} V_{\alpha \beta \epsilon \delta} \Delta {\cal M}_{\alpha \beta , \delta \zeta} -2 i \sum_{\beta \gamma \delta} V_{\zeta \beta \gamma \delta} \Delta {\cal M}_{\epsilon \beta, \gamma \delta},
\end{equation}
and 
\begin{eqnarray}
\label{eqcalm}
 \Delta {\cal M}_{\epsilon \zeta , \mu \nu} & = & i \sum_{\gamma} \left[ t_{\gamma \epsilon} \Delta {\cal M}_{\gamma \zeta, \mu \nu}+t_{\gamma \zeta } \Delta M_{\epsilon \gamma, \mu \nu} - t_{\mu \gamma} \Delta {\cal M}_{\epsilon \zeta, \gamma \nu} - t_{\nu \gamma} \Delta {\cal M}_{\epsilon \zeta , \mu \gamma}\right] \\
& - & 2 i \sum_{\alpha \beta} V_{\alpha \beta  \epsilon \zeta} \left[ \rho_{\alpha \nu} \rho_{\beta \mu} - \rho_{\alpha \mu} \rho_{\beta \nu} + \Delta {\cal M}_{\alpha \beta , \mu \nu}\right] + 2 i \sum_{\gamma \delta} V_{\mu \nu \gamma \delta}\left[ \rho_{\epsilon \delta} \rho_{\zeta \gamma} - \rho_{\epsilon \gamma} \rho_{\zeta \delta} + \Delta {\cal M}_{\epsilon \zeta , \gamma \delta}\right] \nonumber \\
& + & 2 i \sum_{\alpha \beta \gamma} V_{\alpha \beta \gamma \zeta} \left [ 2 \rho_{\alpha \mu} \Delta {\cal M}_{\epsilon \beta, \gamma \nu}+ 2 \rho_{\beta \gamma} \Delta {\cal M}_{\epsilon \alpha, \mu \nu}+ 2 \rho_{\alpha \nu} \Delta {\cal M}_{\epsilon \beta, \mu \gamma} + \rho_{\epsilon \gamma} \left(\Delta {\cal M}_{\alpha \beta , \mu \nu} + \rho_{\alpha \nu} \rho_{\beta \mu} - \rho_{\alpha \mu} \rho_{\beta \nu} \right)\right] \nonumber \\
& + & 2 i \sum_{\alpha \beta \delta} V_{\alpha \beta \epsilon \delta } \left[2 \rho_{\beta \mu} \Delta {\cal M}_{\alpha \zeta , \delta \nu}+2 \rho_{\alpha \delta} \Delta {\cal M}_{\beta \zeta , \mu \nu}+2 \rho_{\beta \nu} \Delta {\cal M}_{\alpha \zeta , \mu \delta} + \rho_{\zeta \delta} \left(\Delta {\cal M}_{\alpha \beta,  \mu \nu} +\rho_{\alpha \nu} \rho_{\beta \mu} - \rho_{\alpha \mu} \rho_{\beta \nu}\right)\right] \nonumber \\
& - & 2 i \sum_{\beta \gamma \delta} V_{\mu \beta \gamma \delta} \left[2 \rho_{\zeta \delta} \Delta {\cal M}_{\epsilon \beta ,  \gamma \nu} + 2 \rho_{\epsilon \delta} \Delta {\cal M}_{\beta \zeta , \gamma \nu} + 2 \rho_{\beta \gamma} \Delta {\cal M}_{\epsilon \zeta \delta \nu} + \rho_{\beta \nu} \left( \Delta {\cal M}_{\epsilon \zeta, \gamma \delta}+\rho_{\epsilon \delta} \rho_{\zeta \gamma}-\rho_{\epsilon \gamma} \rho_{\zeta \delta}\right) \right] \nonumber \\
& - & 2 i \sum_{\alpha \gamma \delta} V_{\alpha \nu \gamma \delta} \left[ 2 \rho_{\zeta \gamma} \Delta {\cal M}_{\epsilon \alpha , \mu \delta} + 2 \rho_{\epsilon \gamma} \Delta {\cal M}_{\alpha \zeta , \mu \delta} + 2 \rho_{\alpha \delta} \Delta {\cal M}_{\epsilon \zeta , \mu \gamma} + \rho_{\alpha \mu} \left( \Delta {\cal M}_{\epsilon \zeta, \gamma \delta} + \rho_{\epsilon \delta} \rho_{\zeta \gamma}-\rho_{\epsilon \gamma} \rho_{\zeta \delta}\right) \right] \nonumber.
\end{eqnarray}
Equations (\ref{eqrho}) and (\ref{eqcalm}) constitute an approximate closed set of equations for the one particle and two particle reduced density matrices which extend TDHF equations to include two-body correlations.  The TDHF equation (\ref{tdhffinal}) can be recovered from Eq. (\ref{eqrho}) by setting $\Delta {\cal M}_{\alpha \beta , \gamma \delta} =0$.

 The exact solution of the Fokker-Planck equation (\ref{FPeq}) does not necessarily satisfy our Gaussian ansatz for $P({\mathbf n})$, which is why Eqs. (\ref{eqrho}) and (\ref{eqcalm}) are approximate. One consequence is that they do not conserve the number of particles. 
This can be seen by calculating the time derivatives of the moments of the number operator
$\hat{N}$, as evaluated using Eqs. (\ref{eqrho}) and (\ref{eqcalm}). We find that the first two moments are  indeed conserved
\begin{equation}
 \frac{d}{d t} \left< \hat{N}\right> = \frac{d}{d t} \left< \hat{N}^2\right>=0.
\end{equation}
However, this is not the case for the third moment 
\begin{equation}
 \frac{d}{d t} \left< \hat{N}^3 \right> \neq 0.
\end{equation}
We conclude that the probability distribution $f(N)$, derived from  Eqs. (\ref{eqrho}) and (\ref{eqcalm}), is not conserved.

It should be emphasized that the approximations leading to the TDHF equations and to the hierarchy derived in this section are of a different nature. The TDHF equations are obtained under the assumption that the state of the system is a single Slater determinant at all times. In contrast, here we assumed that the system can be represented by a Gaussian operator expansion whose probability distribution is Gaussian at all times. This provides a generalization of the TDHF method, suggesting that Eqs. (\ref{eqrho}) and (\ref{eqcalm}) should lead to more accurate results than the TDHF equations. This approximation scheme should be useful for large systems with many particles, which are likely to be less sensitive to changes in the distribution of number of particles.

\section{Unravelling the Fokker-Planck equation in terms of Ito stochastic trajectories}
\label{unraveling}

A probability density evolving according to a Fokker-Planck equation can be computed numerically by simulating an ensemble of trajectories following a stochastic equation of motion~\cite{Breuerbook}. This is commonly used to trade off computer memory by time. This approach was used to study the ground states of Hubbard-type models~\cite{Corney2004,Corney2006b,Assaad2005,Aimi2007,Corboz2008}. Here we present the form of the stochastic equations of motion that naturally connect with the TDHF equations.

Stochastic equations of motion which are equivalent to the Fokker-Planck equation (\ref{FPeq}) are given by
\begin{equation}
 \label{ito}
d n_{\alpha \beta}= {\cal A}_{\alpha \beta} dt + \sum_c \left( {\cal B}^{(1)}_{\alpha \beta, c} d W_c^{(1)} + {\cal B}^{(2)}_{\alpha \beta, c} d W_c^{(2)} \right),
\end{equation}
where we have combined the equations for the real and imaginary parts of $n_{\alpha \beta}$ into a single complex equation.
The noise Wiener increments are real, with Gaussian probability distribution, satisfying $\left< d W_{c}^{(i)} (t) d W _{c^\prime}^{(j)} (t^\prime) \right> = \delta_{cc^\prime} \delta_{ij} \delta \left(t-t^\prime \right)dt$. Equation (\ref{ito}) should be integrated using Ito stochastic calculus~\cite{Gardiner}. The many-body dynamics can be simulated by repeatedly integrating Eq. (\ref{ito}) to create an ensemble of stochastic trajectories. This ensemble, together with Eq. (\ref{rhovian}), or similar identities, can be used to calculate expectation values.

The non-random, or drift, component of the stochastic equation of motion (\ref{ito}), given by (\ref{Aito}), coincides with the right hand side of Eq. (\ref{tdhf3}), as long as we identify $n_{\alpha \beta}$ with $\rho_{\alpha \beta}$. Such identification can naturally be made when the distribution $P({\mathbf n})$ is narrow, or when the noise part is absent (for instance when there are no two-body interactions). 

It is possible to derive an equivalent form of the stochastic differential equations by using a different stochastic calculus~\cite{Gardiner}.  For instance, the Stratonovich calculus results in a similar stochastic equation with ${\cal A}_{\alpha \beta}^{(strat)} \ne {\cal A}_{\alpha \beta}$. This breaks the simple correspondence with TDHF. For completeness, we give an explicit form of the drift terms in the Stratonovich scheme in App.~\ref{stratapp}.

The coefficients ${\cal A}_{\alpha \beta}$, and ${\cal B}_{\alpha \beta,c}^{(1,2)}$, appearing in Eq.({\ref{ito}}), depend quadratically on ${\mathbf n}$. As a result, the nonlinear stochastic trajectories generated from Eq.({\ref{ito}}) may escape to infinity. This is expected to happen in the vicinity of certain phase space regions. This problem may be solved in various ways. When calculating the ground state using an imaginary time equation, one can force the stochastic equations of motion to be real~\cite{Corney2004,Assaad2005,Corney2006b}. This limits the dynamics to a small subset of phase space, which is not likely include the most unstable regions.
Alternatively, some combination of the gauge freedoms, described in App. \ref{gauge}, can be used to obtain an equivalent, more stable, form of the stochastic equations.  A simple systematic way of obtaining stable stochastic equations for the time evolution of general fermionic models is yet to be developed.

\section{Discussion}
\label{disc}

In this paper we have used a Gaussian phase-space representation to generalize a common Hilbert space 
approximation for many-body dynamics, the TDHF, which assumes that the wavefunction of the system is given by a single Slater determinant at all times. 
Here the many body density matrix is represented by an ensemble of Gaussian operators whose distribution of parameters satisfies a Fokker-Planck equation. We have shown that the drift terms of this Fokker-Planck equation have similar form to the TDHF equations [compare Eqs. (\ref{Aito}) and (\ref{tdhf3})]. This suggests that the phase-space representation can serve as a good basis for new approximation schemes, which systematically go beyond TDHF.

The Fokker Planck equation (\ref{FPeq}) maps the original fermion system rigorously into a equivalent set of classical oscillators $n_{\alpha \beta}$ with stochastic dynamics.  The number of oscillators is the same as the number of elements of the reduced single electron density matrix and it thus scales as $\sim N^2$ with the basis set size.  The TDHF is a mean field approximation to this stochastic dynamics whereby the same classical coordinates satisfy a deterministic equation.  Generalizations of the type given here were conjectured in ref.~\onlinecite{Tretiak1998} but the GPSR puts it on a firm theoretical basis.  

Linear and nonlinear response functions e.g. optical spectra have a very different formal structure for quantum and classical systems.  To calculate the response function to a classical field $E(t)$ one needs to add a coupling term $E(t) A$ to the Hamiltonian where $A$ is a dynamical variable.  $n$’th order quantum response functions are then given by $n$ nested commutations such as $\left< \left[ \left[ A(\tau_1), A(\tau_2)\right],A(\tau_3)\right] \right>$
 which gives $2^n$ terms (Liouville space pathways).  Classical response functions have a very different form and require to run groups of several nearby trajectories and carefully monitor how they diverge.  The responce functions can be expressed in terms of the stability matrices of the stochastic trajectories of the free system (without external driving)~\cite{Mukamel1996,Dellago2003}, or by solving the distribution of the driven system and combining terms with different orders of interactions (finite field techniques)~\cite{Mukamel2004}. An intriguing aspect of the present approach is that the quantum response functions are recast rigorously in a classical form.  This could result in new insights and could suggest new numerical simulation techniques for fermions.

Studying the merits of this new representation will be an interesting future direction. Moreover, the quantum response function is one specific combination of the Liouville space pathways.  Other combinations represent spontaneous fluctuations and how they are affected by external driving~\cite{Harbola2004,Cohen2003}.  Many attempts have been made to connect nonlinear response and fluctuations through generalized fluctuation-dissipation relations~\cite{Wang2002}.  The GPSR may shed a new light into this issue and provide a classical picture for the connection between quantum response and fluctuations.

We have constructed an approximate, simple, numerical scheme
by further assuming that the probability distribution $P({\mathbf n})$
has a Gaussian form at all times. The hierarchy of equations for reduced many-body density matrices can be truncated at the single and two particle level. This yields a coupled set of equations, (\ref{eqrho}) and (\ref{eqcalm}), which can be solved to calculate the time dependence of expectation values of all the one and two-body operators.

The exact time evolution of the phase-space distribution function, which represents the many body density matrix, conserves the distribution of the number of electrons. This is not true for the approximate dynamics of Eqs. (\ref{eqrho}) and (\ref{eqcalm}). While the mean number of electrons, and its variance, do not vary with time, higher moments are not conserved. This approximation scheme thus suffers from the old $N$ representability problem~\cite{Coleman} common to other reduced density matrix approaches. Nevertheless, this scheme may be a useful description, most likely for systems with many electrons, where expectation values are less sensitive to small variations in the distribution of the number of particles.

The phase-space representation allows the development of new approximation schemes, by making other assumptions on the form of $P({\mathbf n})$. It may be possible to include constraints that ensure that approximations on $P({\mathbf n})$ conserve the distribution of number of particles, while still allowing to truncate the hierarchy. More work is needed in order to gain understanding on the way in which various assumptions on $P({\mathbf n})$ manifest themeselves on the many body dynamics.

\section*{Acknowledgments}

We gratefully acknowledge the support of the National Science Foundation through grant No. CHE-0745892.

\begin{appendix}

\section{Gauge freedom in the Fokker-Planck equation}
\label{gauge}

As is the case for Bosonic coherent states~\cite{Glauberbook}, the Gaussian operator basis for fermions is overcomplete. This means that there exist an infinite number of equivalent forms of the time evolution equations for $P({\mathbf n})$ which give the same expectation values of all physical observables. Below we give a brief description of the various ways of obtaining such equivalent representations, following Ref.~\onlinecite{Corney2006b}, and examine whether they retain the correspondence between the drift terms and the TDHF equations.

A  way of changing the Fokker-Planck equation, without affecting the physical observables, is to add formally vanishing operators, such as $\hat{c}_{\alpha}^{\dagger} \hat{c}_{\alpha}^\dagger \hat{c}_\beta \hat{c}_\gamma$, to the Hamiltonian, and then use Eqs (\ref{ident1}) and (\ref{ident2}) to replace these terms by a differential operator. This method, which is only applicable for systems of identical fermions, was termed Fermi gauge in Refs.~\onlinecite{Corney2004, Corney2006b}. A direct calculation shows that such Fermi gauges do not change the form of the drift terms ${\cal A}_{\alpha \beta}$ in Eq.(\ref{FPeq}), but do affect the noise terms. Fermi gauges were used to obtain imaginary-time equations for $P$ where all the parameters are real rather than complex valued.

A different method of obtaining equivalent forms of the equations of motion is termed drift gauge~\cite{Deuar2002,Corney2006b}. Here one adds a new parameter, which corresponds to a weight given to each stochastic trajectory, thus replacing the (uniform) average over trajectories by a weighted average. As a result, a new distribution function is defined in a larger parameter space, which includes the weight as a variable, and a Fokker-Planck equation is derived for this new distribution functions. At the same time, the evaluation of averages using the distribution $P$ is replaced by weighted averages involving the new weight parameter.
It is then possible to modify the drift terms ${\cal A}_{\alpha \beta}$, while at the same time adding suitable noise terms to part of the Fokker-Planck equation related to the new weight parameter, in such a way that will not change the values of weighted averages~\cite{Corney2006b,Assaad2005}. Drift gauges have been used to change the drift terms in order to avoid trajectories which run to infinity. 

An equivalent, but different form of the stochastic equations of motion can be obtained by using different forms for the decomposition of ${\cal Q}$ in Eq. (\ref{defQ}). This will not affect Eq.(\ref{FPeq}), but will change its stochastic trajectory simulation, for instance using Eq. (\ref{ito}). However, when combined with drift gauges the choice of solution of Eq. (\ref{defQ}) may result in a different Fokker-Planck equation in the generalized phase space. The freedom to choose different solutions of Eq. (\ref{defQ}) was termed diffusion gauge in Refs.~\onlinecite{Deuar2002,Assaad2005}.

The gauge freedoms may be useful for practical implementation of the GSPR. If a solution of Eq.~(\ref{FPeq}) leads to a probability distribution with long tails, and therefore  for the appearance of boundary terms, due to the integration by parts,
the gauge freedom can be used to generate an equivalent representation whose probability distribution is more localized. One can hope to use the many ways to generate equivalent representations, by combining the various gauges, to find a representation which does not suffer from boundary terms.

\section{The Stratonovich stochastic equations}
\label{stratapp}

The Stratonovich calculus is an alternative to the Ito calculus used in eq. (\ref{ito}). In this scheme the terms in the stochastic differential equation are evaluated at the midpoint of the interval, as opposed to the initial point used in the Ito scheme.

In practice, one solves the finite differences equation
\begin{equation}
 {n}_{\alpha \beta}^{(i+1)}-n_{\alpha \beta}^{(i)}= {\cal A}_{\alpha \beta}^{(strat)} \left( {\mathbf n}^{mid} \right) dt+\sum_{c} \left( {\cal B}^{(1)}_{\alpha \beta, c} ({\mathbf n}^{mid}) d W_c^{(1)} + {\cal B}^{(2)}_{\alpha \beta, c}({\mathbf n}^{mid}) d W_c^{(2)} \right),
\end{equation}
with ${\mathbf n}^{mid} \equiv ({\mathbf n}^{(i+1)}+{\mathbf n}^{(i)})/2$, in order to get ${\mathbf n}^{(i+1)}$. This implicit equation can be solved by iteration. It is of interest to note that implicit methods tend to show better numerical stability then explicit ones, see Ref.~\onlinecite{Drummond1991} for a review.

This stochastic integration should lead to the Fokker-Planck equation (\ref{FPeq}). However, this means that the coefficients ${\cal A}^{(strat)}$ can not be equal to the ones appearing in Eq. (\ref{Aito}). This is due to correlations in the noise terms, which lead to the second order derivatives with the form $\frac{1}{2} \frac{\partial}{\partial n} {\cal B} \frac{\partial}{\partial n} {\cal B} P$, rather than $\frac{1}{2} \frac{\partial}{\partial n}\frac{\partial}{\partial n} {\cal B} {\cal B}P$ which would appeared when using the Ito scheme. (We suppressed subscripts for brevity.) 

To compensate for these correlations, the drift terms of the Ito and Stratonovich schemes are related by
\begin{equation}
 {\cal A}_{\alpha \beta}^{(strat)} = {\cal A}_{\alpha \beta} - \frac{1}{2} \sum_{\gamma \delta} \sum_{c} {\cal B}_{\gamma \delta,c}^{(1)} \frac{\partial}{\partial n_{\gamma \delta}} {\cal B}_{\alpha \beta ,c}^{(1)} - \frac{1}{2} \sum_{\gamma \delta} \sum_{c} {\cal B}_{\gamma \delta,c}^{(2)} \frac{\partial}{\partial n_{\gamma \delta}} {\cal B}_{\alpha \beta ,c}^{(2)}.
\end{equation}
For the Hamiltonian (\ref{universalh}) we find
\begin{eqnarray}
 {\cal A}_{\alpha \beta}^{(strat)} & = &-i \sum_{\gamma} \left( t_{\beta \gamma} n_{\alpha \gamma}-t_{\gamma \alpha} n_{\gamma \beta}\right) -i \sum_{\gamma \delta} \left( V_{\beta \delta \delta \gamma} n_{\alpha \gamma}-V_{\gamma \delta \delta \alpha} n_{\gamma \beta}\right) \nonumber \\
& + & i\sum_{\gamma \delta \epsilon}  n_{\gamma \delta} n_{\alpha \epsilon} \left(V_{\beta \gamma \epsilon \delta}+ V_{\gamma \beta \delta \epsilon} \right)-i \sum_{\gamma \delta \epsilon} n_{\gamma \delta} n_{\epsilon \beta} \left( V_{\epsilon \gamma \alpha \delta}+V_{\gamma \epsilon \delta \alpha}\right). 
\end{eqnarray}
We have just shown that the Stratonovich form of the drift term differ from the one obtained using the Ito scheme, and therefore,  differs from the terms appearing in the TDHF equations of motion. Nevertheless, by construction, both stochastic schemes give the same ensemble of trajectories.

\end{appendix}

\end{document}